\documentclass[aps,prb,twocolumn,showpacs,amsmath,amssymb]{revtex4}
\usepackage{graphicx}
\usepackage{epstopdf}

\begin{document}
\title{Metal-insulator transition and magnetism in correlated band insulator: FeSi and Fe$_{1-x}$Co$_{x}$Si}
\author{V.V. Mazurenko,$^{1}$ A.O. Shorikov,$^{1,2}$ A.V. Lukoyanov,$^{1,2}$ K. Kharlov,$^{1}$ E. Gorelov,$^{3}$ A.I. Lichtenstein,$^{4}$ V.I. Anisimov$^{1,2}$}
\affiliation{$^{1}$Theoretical Physics and Applied Mathematics Department, Urals State Technical University, Mira Street 19,  620002
Ekaterinburg, Russia \\
$^{2}$Institute of Metal Physics, Russian Academy of Sciences, 620219 Ekaterinburg GSP-170, Russia \\
$^{3}$Institut f$\ddot{u}$r Festk$\ddot{o}$rperforschung and Institute for Advanced Simulation, Forschungzentrum J$\ddot{u}$lich, 52425 J$\ddot{u}$lich, Germany  \\
$^{4}$Institute of Theoretical Physics, University of Hamburg, Jungiusstrasse 9, 20355 Hamburg, Germany }
\date{\today}

\begin{abstract}
The LDA+DMFT (local density approximation combined with dynamical mean-field theory)
computation scheme has been used to study spectral and magnetic properties 
of  FeSi and  Fe$_{1-x}$Co$_{x}$Si. 
Having compared different models we conclude that a correlated band insulator scenario in contrast to Kondo insulator model agrees with FeSi band structure  as well as experimental data. Coulomb correlation effects lead to band narrowing  of the states near the Fermi level with mass renormalization parameter $m^*\approx 2$ in agreement with the results of angle-resolved photoemission
spectroscopy (ARPES). Temperature dependence of spectral functions and magnetic susceptibility calculated in DMFT reproduces transition from nonmagnetic semiconductor to metal with local magnetic moments observed experimentally. Cobalt doping leads to ferromagnetism that has itinerant nature and can be successfully described by  LDA+DMFT method.
\end{abstract}

\maketitle

\section{Introduction}
\label{intro}
The narrow-gap semiconductor FeSi demonstrates an interesting interplay between magnetic and electronic properties. Magnetic susceptibility temperature dependence shows maximum at 500 K and Curie-Wiess behavior at higher temperatures. \cite{jaccarino} While  intrinsic magnetic
susceptibility vanishes below 50 K,  FeSi does not show any sign of spin ordering down to the lowest temperatures. \cite{watanabe} Photoemission \cite{ishizaka} and optical experiments show an energy gap of about 60 meV at low temperatures that is gradually filled with temperature increase. \cite{optics} A resistivity temperature dependence shows transition from a narrow gap semiconductor to a bad metal. \cite{rho} Cobalt doping results in ferromagnetic metal state for Fe$_{1-x}$Co$_{x}$Si. \cite{Beille2}

Several models have been suggested to explain the
unusual temperature dependence of the FeSi physical properties ranging from spin fluctuations \cite{takahashi} to phenomenological
models assuming two narrow $d$-bands in the
vicinity of the band gap. \cite{mandrus,Fu} Such density of
states (DOS) models are similar to a Kondo insulator description,
and due to the striking similarities in the physical
properties of these Kondo insulators it was claimed that
FeSi is the first Kondo insulator containing no $f$-electrons. \cite{fath,fisk1}

Historically, the first FeSi model was proposed by Jaccarino et al. \cite{jaccarino} To describe the unusual magnetic susceptibility and specific heat a  model DOS was proposed with extreme narrow band peaks around a small energy gap.  Despite the fact that  the model fitting results were in  good agreement with experimental data for the susceptibility this picture contradicts a band structure calculation  \cite{mattheiss}  where no unphysical narrow bands were obtained.

The next model proposed to describe electronic and magnetic properties of FeSi was the Kondo insulator model.\cite{mason}  The motivation of applying the Kondo model to explain physical properties of FeSi was that the spin-fluctuation spectra of CeNiSn and FeSi are similar. According to this model, a set of localized atomic-like electron levels interact with a wide itinerant band. The insulating state scaled by a Kondo temperature T$_K$ is the result of a weak hybridization between the localized and itinerant bands. The implementation of the Kondo insulator model to FeSi compound is however questionable since the band structure of FeSi shows a strong hybridization between the Fe-3$d$ and Si-$p$ states. Also the magnetic interactions in FeSi are essentially ferromagnetic and not antiferromagnetic, as it would be expected from an RKKY-picture. For all these reasons it is desirable to have a microscopic model based on the realistic band structure which can reproduce experimental data.  

First-principle band structure calculations of Mattheis and Hammann \cite{mattheiss} have shown that energy gap value of 0.1 eV is essentially smaller than the width of the band above the Fermi level ($\sim$ 0.5 eV).  Band structures analysis led  the authors to conclusion that hybridization between Fe-$d$ and Si-$p$ states is very strong. These results do not support the Jaccarino's model as well as the Kondo scenario.

Recently, another model of a correlated band insulator (CBIM) was proposed by Kunes and Anisimov.\cite{Kunes}  Using the Dynamical Mean-Field Theory the authors have taken into account local dynamical correlations for a gapped LDA spectral function of FeSb$_2$ that experimentally demonstrates transition with temperature increase from nonmagnetic semiconductor to metal with local moments similar to FeSi. It was found that the DMFT energy gap is reduced due to correlation effects by a factor of two from its LDA value. Within the CBIM picture bands above and below the energy gap are formed by nonlocal bonding and antibonding orbital combinations. The on-site Coulomb interaction leads to a competition between localization and formation of nonlocal bonds. The DMFT calculations allowed to reproduce successfully  temperature dependence of magnetic susceptibility, resistivity and optical conductivity experimentally observed in FeSb$_2$.

Recent  angle-resolved photoemission experiments \cite{Klein} have proved validity of the correlated  band insulator scenario for FeSi. The authors have found an effective band-mass renormalization m$^*$/m$\approx$2. As a consequence the LDA band gap of 100 meV is renormalized  by a factor of 0.5 to about 60 meV in agreement with other experimental data.

 We summarized the parameter values of the models used to describe FeSi  in Table \ref{classification}. The basic difference between Kondo insulator and correlated band insulator (CBIM) models is a ratio between energy gap value E$_{gap}$ and a width $W$ of the bands around the gap. While Kondo insulator model requires $W/E_{gap}<<$1 to give good results, CBIM uses $W/E_{gap}>>$1 in agreement with FeSi realistic band structure. 

\begin{table}[ht]
\centering
\caption{Classification of FeSi models. }
\label{classification}
\begin {tabular}{cc}
  \hline
  Jaccarino's model \cite{jaccarino}               & W $<<$ E$_{gap}$  (W $\rightarrow$ 0 K, E$_{gap}$ = 1520 K) \\
  Kondo \\
  insulator model  \cite{mandrus}                                    &  W = E$_{gap}$/2    (W = 500 K, E$_{gap}$ = 1000 K) \\
  CBIM (this work)   &  W $>>$ E$_{gap}$    (W = 5000 K, E$_{gap}$ = 1000 K   )    \\
  \hline
\end {tabular}
\end {table}

This paper is aimed to provide microscopic analysis of hybridization and correlation processes in FeSi.
First we propose a simple band insulator model that captures the essential hybridization effects between Fe-3$d$  and Si-3$p$ states in FeSi (Section \ref{FeSi-band}).  This model demonstrates an energy gap that is five times narrower than the width of bands. In order to investigate correlation effects  we define a density of states model based on LDA calculations for FeSi and solve it using the Dynamical Mean-Field Theory (Section \ref{corr-effects}). A strong renormalization of DOS near the Fermi level was found in the DMFT calculations that is in good agreement with a band narrowing observed in the recent ARPES experiments.  Temperature increase in DMFT calculations results in transition from nonmagnetic insulator to a bad metal with local moments in agreement with experimental data.
Having encouraged by these results we have investigated a doped FeSi model to study  the magnetic properties of Fe$_{1-x}$Co$_{x}$Si alloys (Section \ref{FeCo}). We have demonstrated that a good agreement between calculated and experimental values for magnetization $M$ as a function of doping $x$ can be achieved using the LDA+DMFT method.

\section {${\bf FeSi}$}
\subsection{Band effects} 
\label{FeSi-band}
The crystal structure of FeSi corresponds to four formula units in unit cell and its band structure is rather complicated for analysis using a simple model.   
To provide a better understanding of FeSi band structure near the Fermi level Mattheiss and Hamann \cite{mattheiss}
have proposed to consider the closely related phase with the symmetry of the rocksalt structure. The latter phase can be obtained from original crystal structure by relatively small shift of atomic positions. They have shown that the origin of the FeSi energy gap can be traced to a pseudogap that is present in the rocksalt phase. This result is a starting point of our investigation that is aimed to construct a minimal realistic band insulator model which captures main features of LDA electronic spectrum of FeSi. 
\begin {figure}[!h]
\includegraphics[width=0.42\textwidth]{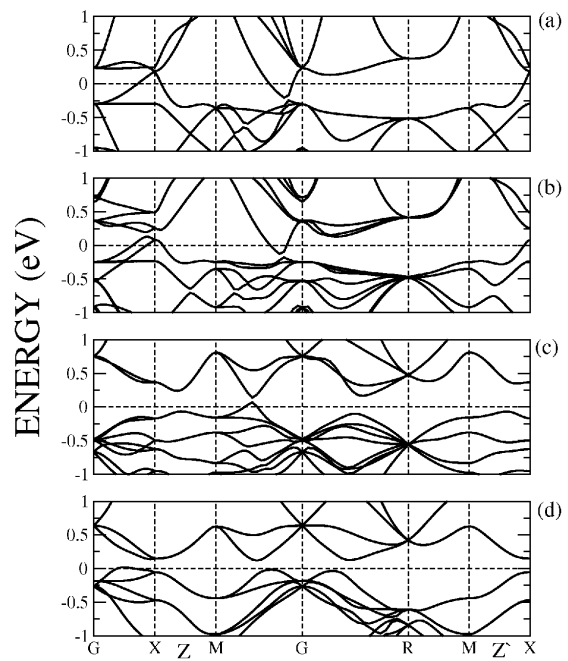}
\caption { TB-LMTO energy band of FeSi obtained with different sets of atomic position
parameters u(Fe) and u(Si). (a) Structure with u(Fe)=0.25 and u(Si)=0.75 corresponds to
nonprimitive rocksalt structure that contains four FeSi formula units. Figures (b) and (c) are energy band results 
for two transitional structures with u(Fe)=0.23,u(Si)=0.76 and u(Fe)=0.19,u(Si)=0.79, respectively.
Figure (d) corresponds to the real FeSi structure. Symmetry lines are chosen according to Ref. \onlinecite{mattheiss}}
\label{bands-u}
\end{figure}

As first step we have studied effects of atomic positions shift on the band picture of FeSi. 
For that purpose we have carried out 
calculations using the Tight Binding Linear-Muffin-Tin-Orbital 
Atomic Sphere Approximation (TB-LMTO-ASA) method with conventional local-density approximation (LDA) \cite{Andersen} 
for real (simple cubic) structure (atomic positions u(Fe)=0.1358 and u(Si)=0.844), rocksalt (face-centered-cubic) phase (u(Fe)=0.25 and u(Si)=0.75)
and for two model  structures with atomic positions  intermediate between those values (Fig. \ref{bands-u}). The obtained bands agree with LAPW results for real and rocksalt phases.\cite{mattheiss} 
The fcc band structure contains several electrons and holes pockets near $G$ and $X$ points in the Brillouin zone that corresponds to a pseudogap in the energy spectrum in contrast to the real gap in the simple cubic structure. For intermediate structure with atomic position parameters u(Fe)=0.19 and u(Si)=0.79, the energy gap is open along all symmetry lines of the Brillouin zone with the exception of $GM$. One can see that there is a complete energy gap in the real FeSi structure.  This result agrees with a conclusion of Mattheiss and Hamann \cite{mattheiss} that the energy gap state of FeSi results from the distortion of the pseudogapped rocksalt structure. 

\begin {figure}[ht]
\begin{center}
\vspace{1cm}
\includegraphics[width=0.4\textwidth]{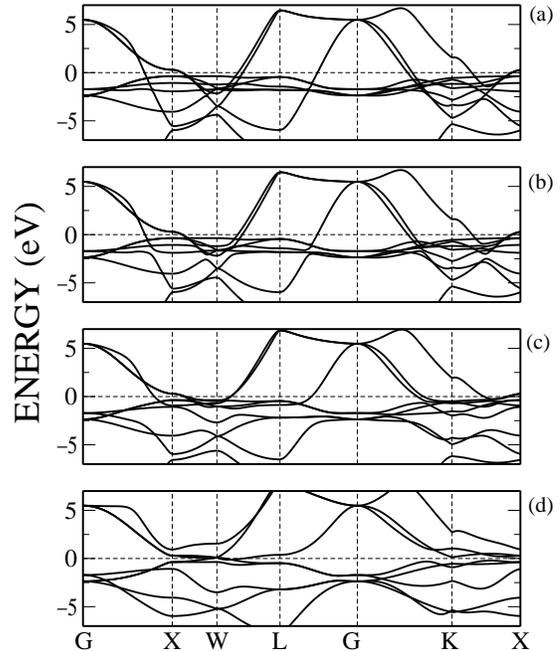}
\caption { Band structures for the different degree of the hybridization strength between
Fe-3$d$ and Si-3$s$, 3$p$ states in fcc crystal structure with one formula unit in unit cell.
Figures (a), (b), (c) and (d) are the energy spectra for hybridization scaling parameter $\alpha$=0; 0.1; 0.5 and 1, respectively.
Symmetry lines are chosen according to Ref. \onlinecite{mattheiss}.}
\label{bands-a}
\end{center}
\end{figure}

It is naturally to expect that the features of the FeSi electronic spectrum (such as the narrow gap and the peak above the Fermi level) are provided by Fe-3$d$ and Si-3$s$, 3$p$ states. \cite{mattheiss,anisimov} To investigate the origin of the pseudogap state in the fcc phase we artificially scaled the hybridization strength between Si-3$s$, 3$p$ and Fe-3$d$ states. To do so off-diagonal Hamiltonian elements which describe the hybridization between Fe and Si atoms were multiplied by a coefficient $0\leq\alpha\leq 1$.  These results are presented in Fig. \ref{bands-a}.

To simplify our analysis of the band structures one can consider dispersion curves along $LG$ symmetry line  where complete energy gap opens due to hybridization between Fe-3$d$ and Si-3$s$, 3$p$ states.
Without hybridization ($\alpha=0$) the wide silicon band 
crosses three narrow 3$d$-bands of iron  for $k$-vectors along $LG$ direction. At increasing hybridization strength (Fig. \ref{bands-a}b and \ref{bands-a}c) two of the iron bands strongly interact with silicon states while one of them remains practically unaffected by $d-p$ hybridization. 
The resulting dispersion curves of the fully hybridized system are presented in Fig. \ref{bands-a} (d).
It is clear that the pseudogapped state (Fig. \ref{bands-a}d)  originates from a strong hybridization 
of the wide silicon band ($\approx$ 10 eV) with the relatively narrow iron band ($\approx$ 1 eV). The narrow gap opens between antibonding hybridized band above and unhybridized band below the Fermi level.

While qualitatively this situation resembles the Kondo insulator picture with weakly hybridized wide and narrow bands, there are two essential quantitative differences. First of all  the ``narrow'' band having width of an order of magnitude smaller than the ``wide'' band is still too wide in absolute value of about 1 eV. At second $d-p$ hybridization is so strong that ``pure hybridization'' gap in \mbox{Fig. \ref{bands-a}d} is larger than 2 eV. The small value of the gap is not due to $d-p$ hybridization weakness but happens between hybridized  and unhybridized bands.

\begin {figure}[t]
\includegraphics[width=0.4\textwidth]{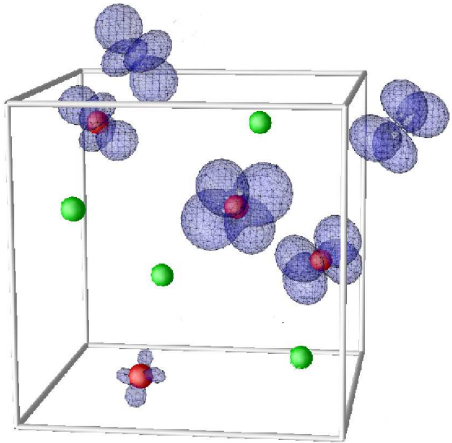}
\caption {(Color online) Wannier function corresponding to a narrow band above the Fermi level (see Fig. \ref{bands-u}d). The Wannier function is centered at a 3$d$ orbital of iron atom (red spheres).  We found about 40\% of the electron density at the central atom and its iron neighbours. Green spheres correspond to silicon atoms.}
\label{Wannier}
\end{figure}

Another argument against the Kondo scenario can be found from analysis of the two bands  (Fig. \ref{bands-u}d) forming a well-separated narrow peak above the Fermi level (Fig. \ref{LDA-dos}). In Kondo system these bands should be associated with strongly localized atomic orbitals of an iron atom. To check this we have calculated  Wannier functions for two  bands above the Fermi level using a projection procedure. \cite{proj} A spatial distribution for one of the calculated Wannier functions is shown in Fig. \ref{Wannier}. The resulting Wannier functions correspond to a complex combination of 3$d$-states of iron and 3$s$-, 3$p$-states of silicon and are spread over whole unit cell of FeSi containing eight atoms. This picture is very far from localized atomic orbital needed for the Kondo scenario and supports the band insulator model.

\begin{figure}[hb]
\includegraphics[width=0.5\textwidth]{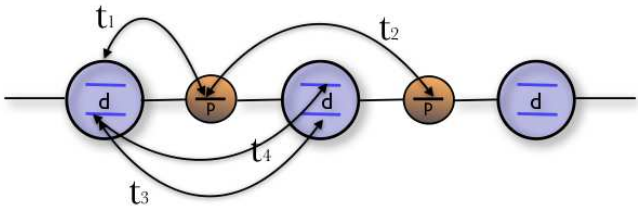}
\caption {(Color online) One-dimensional effective model. Large and small circles correspond to iron and silicon atoms in the fcc phase of FeSi.}
\label{one-d}
\end{figure}

We built an effective microscopic model (Fig. \ref{one-d}) for $LG$ direction, that contains a minimal set of orbitals 
and reproduces the gapped state. The model silicon and model iron are described with one and two orbitals, respectively.
The model Hamiltonian is given by
\begin{displaymath}
H= \left( \begin{array}{ccc}
 2t_{3}cos(k) & 2t_{4}cos(k) & 2t_{1}cos(k/2) \\
2t_{4}cos(k) & 2t_{3}cos(k) & 2t_{1}cos(k/2) \\
2t_{1}cos(k/2) & 2t_{1}cos(k/2) &2t_{2}cos(k)
\end{array} \right),
\end{displaymath}
where $t_1$, $t_2$, $t_3$ and $t_4$ are hoppings between model iron and silicon orbitals presented in Fig.\ref{one-d}. 
We have estimated the hopping parameters of the $LG$ model using the real band structure of the full Hamiltonian presented in Fig. \ref{bands-a}.
The obtained hopping integrals are t$_{1}$=2.0 eV, t$_{2}$=3.0 eV, t$_{3}$=-1.2 eV, t$_{4}$=-0.25 eV.
These values are much larger than those calculated for localized systems \cite{mazurenko} and are far outside of the values range needed for the Kondo insulator scenario.

\begin {figure}[t]
\includegraphics[angle=270,width=0.4\textwidth]{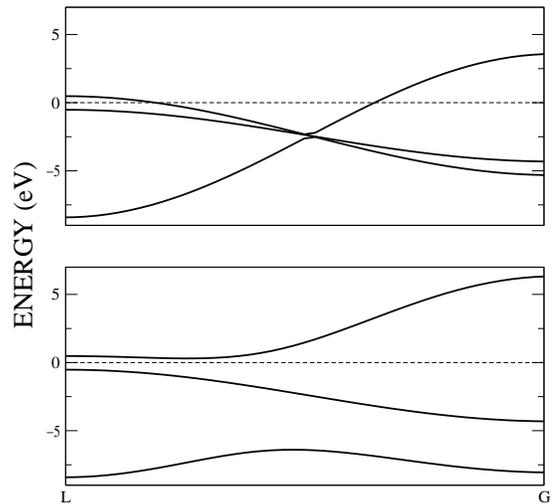}
\caption {Model band structures. Lower and upper figures correspond to the models with and without hybridization, respectively.}
\label{model-t}
\end{figure}

The calculated model band structures with ($\alpha$=1) and without ($\alpha$=0) hybridization are presented in Fig. \ref{model-t}.
 One can see that our model results are in good agreement with those obtained from calculations with full Hamiltonian (Fig. \ref{bands-a}). 
 The substantial Fe-Si hybridization leads to splitting into a lower bonding and an upper antibonding bands with a non-bonding band in between. 
We have obtained a small gap semiconducting ground state with the energy gap value of 0.8 eV that is much smaller than the narrowest  band width of 4.8 eV near the Fermi level. 
One should note that the small value of the energy gap is not a consequence of a weak Fe-Si hybridization and the minimal realistic band structure model of FeSi differs qualitatively from the Kondo insulator regime.

The proposed band model can be used to study Coulomb correlation problem within statical or dynamical mean-field approaches. However, as we will show below, it is possible to define an effective density of states model derived from the full DOS obtained in LDA calculation for the real crystal structure. Solution of this model by DMFT allows us to describe anomalous physical  properties of FeSi.

\subsection{Correlation effects}
\label{corr-effects}
\begin{figure}[ht]
\includegraphics[width=0.40\textwidth, angle=0]{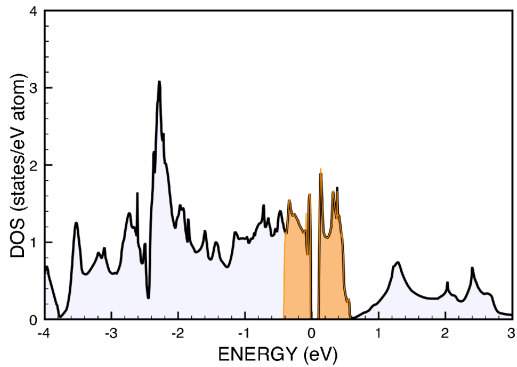}
\caption {(Color online) Model (filled area) and full (line) densities of states.}
\label{LDA-dos}
\end{figure}

In this section we study the effects of the Coulomb correlation on the electronic structure and physical  properties of FeSi using the Dynamical Mean-Field Theory.  
As an input DMFT requires a non-interacted Hamiltonian or a density of states (DOS).\cite{DMFT} The essential features of FeSi density of states obtained in LDA calculations (Fig. \ref{LDA-dos}) are small energy gap $\approx$100 meV and narrow ($\approx$0.5 eV) peak above the  Fermi level containing 0.5 electrons per spin per Fe atom. We have defined  the model density of states (see filled area in Fig. \ref{LDA-dos}) by cutting from the entire DOS area around Fermi energy with an integral equal to 1 electron  per spin per Fe atom. 
The model density of states contains the main features of the FeSi spectrum, namely  the energy gap (0.1 eV) and the narrow peak above  the Fermi level (0.5 eV). 
 
As impurity solvers of the DMFT problem we have used a Quantum Monte-Carlo method with Hirsch-Fye algorithm \cite{HF} (QMC-HF), a Continuous-time Quantum Monte-Carlo method with interaction expansion \cite{CT-QMC} (CT-QMC) and an exact diagonalization temperature dependent Lanczos \cite{Capone} approach. 

The value of the on-site Coulomb interaction parameter $U$ was chosen to be 1 eV that is close to the value used in Ref. \onlinecite{Kunes}. This
rather small value can be justified by an effective screening of $d-d$ Coulomb interaction due to strong Fe-ligand hybridization as it was demonstrated in constrain $DFT$ calculation of the Coulomb interaction parameter $U$ for LaOFeAs.\cite{laofeas}

\begin{figure}[!b]
\centering
\includegraphics[width=0.40\textwidth, angle=0]{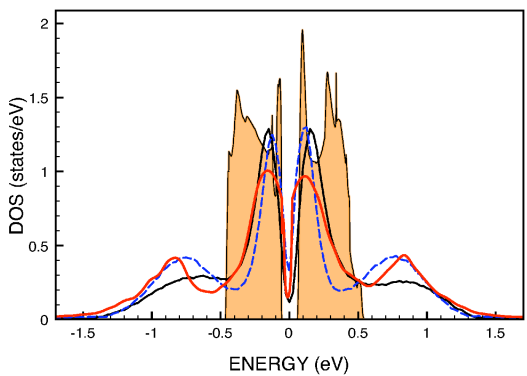}
\caption {(Color online) Spectral functions obtained from DMFT calculations using QMC-HF (blue dashed line), CT-QMC (black solid line) and exact diagonalization (red solid line) techniques at T = 232 K. Orange filled area corresponds to LDA density of states.}
\label{dmft-dos}
\end{figure}

The resulting paramagnetic densities of states at T = 232 K calculated in DMFT using various impurity solvers are presented in Fig. \ref{dmft-dos}.
One can see that all methods result in a pseudogapped state with strongly renormalized density of states near the Fermi level.
There are satellites at $\pm U/2$ which correspond to the lower and upper Hubbard bands.

The energy area around the Fermi level corresponds to quasiparticle states that are usually described as non-interacting bands renormalized by Coulomb correlations.
The renormalization process can be understood by a low-frequency analysis of the dynamical mean-field equations. 
The lattice Green function is given by
\begin{eqnarray}
\label{Green}
G(\omega) = \sum_{{\bf k}} (\omega - \Sigma (\omega) - \epsilon ({\bf k}) )^{-1},
\end{eqnarray}
where $\Sigma$ is a self-energy and $\epsilon ({\bf k})$ is a LDA spectrum.
We expand the real part of the self-energy in the vicinity of the Fermi energy leaving only linear term:
\begin{eqnarray}
Re \Sigma (\omega) \approx Re \Sigma (0) + \omega \frac{d Re \Sigma (\omega)}{ d \omega} |_{\omega = 0}.
\end{eqnarray}
Then Green function for specific wave vector ${\bf k}$ is
\begin{eqnarray}
\label{G-m}
G_{{\bf k}}(\omega) \approx (\omega m^* - \epsilon({\bf k}))^{-1},
\end{eqnarray} 
where $m^*$ is the effective band mass renormalization parameter
\begin{eqnarray}
m^*\equiv 1-\frac{d Re \Sigma (\omega)}{ d \omega} |_{\omega = 0} .
\label{mstar}
\end{eqnarray}
Equation (\ref{G-m})  can be rewritten as
\begin{eqnarray}
\label{Z}
G_{{\bf k}}(\omega) \approx \frac{Z}{\omega - \tilde \epsilon({\bf k})}, 
\end{eqnarray}
where $Z\equiv 1/m^*$ is a quasiparticle weight and $\tilde \epsilon ({\bf k}) \equiv \epsilon ({\bf k})/m^*$ is a new band dispersion.
Therefore, the Coulomb correlation renormalization for quasiparticle states near the Fermi level results in a band structure narrowing by a factor of $m^*$ and a reduction of the corresponding spectral weight  by a factor of $Z$ with the rest of the spectral weight transfered to upper and  lower Hubbard bands at $\pm U/2$.

In our DMFT calculations the effective band mass renormalization parameter $m^*\approx 2$ was found in good agreement with the results of angle-resolved photoemission
spectroscopy (ARPES) \cite{Klein} showing band narrowing by a factor of two in comparison with bands calculated in LDA.

\begin{figure}[!t]
\includegraphics[width=0.40\textwidth, angle=0]{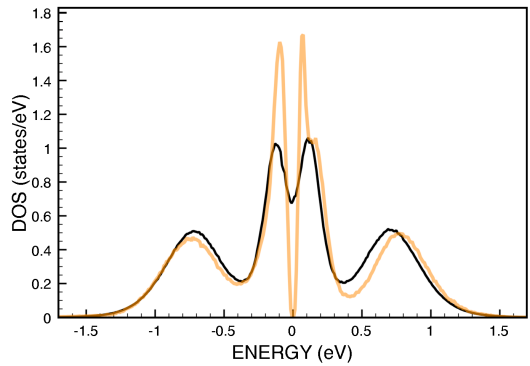}
\caption {(Color online) Density of states obtained from QMC calculations at T= 386 K (black line) and 96 K (orange line).}
\label{dos-T-QMC}
\end{figure}

Experimentally FeSi demonstrates transition with temperature increase from narrow gap semiconductor to bad metal. \cite{rho} 
We have performed DMFT calculations at different temperatures and such an experimentally observed transition was successfully reproduced theoretically.
The DMFT spectral functions calculated for different temperatures are presented in Fig. \ref{dos-T-QMC}.
The energy gap of about 50 meV found at $T$= 96 K agrees with experimental results of resistivity measurements which indicate a charge gap of about 60 meV.  \cite{rho1}  At $T$= 386 K the gap area is nearly completely filled by spectral weight transfer from sharp peaks near the gap resulting in the spectral function corresponding to a bad metal. The energy gap value obtained in the DMFT calculation is two times smaller than corresponding value from LDA band structure calculations (0.1 eV). That agrees very well with the effective band mass renormalization parameter $m^*\approx 2$ Eq.\eqref{mstar} obtained in our DMFT calculations.
\begin{figure}[!t]
\centering
\includegraphics[width=0.40\textwidth, angle=0]{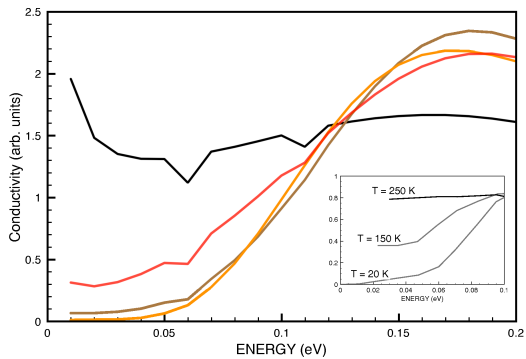}
\caption {(Color online) The convolution for FeSi model at 386 K (black line), 232 K (red line), 116 K (brown line) and 96 K (orange line). Inset: The experimentally observed optical conductivity taken from Ref. \onlinecite{optics} }
\label{optics}
\end{figure} 

Optical spectroscopy experiments show gradual filling of the low-temperature energy gap with temperature increase till complete gap disappearance  above room temperature. \cite{optics} We have estimated the optical conductivity by a spectral function convolution using the following expression
\begin{eqnarray}
\sigma (\omega, T) = \frac{1}{\omega}\int d\epsilon N(\epsilon) N (\epsilon + \omega) [1-f(\epsilon+\omega,T)],
\end{eqnarray}
where $N (\epsilon)$ is a spectral function obtained from DMFT calculations (Fig. \ref{dmft-dos}) and $f(\epsilon, T)$  is the Fermi distribution function. The calculated convolution together with experimental data for temperature dependent  optical conductivity \cite{optics} are presented in Fig. \ref{optics}.

At low temperatures a well pronounced  energy gap of about 0.05 eV can be observed in both experimental and theoretical curves. With temperature increase optical conductivity increases at the energies below 0.05 eV and finally at T= 386 K there is no any trace of the gap in the theoretical curve in  good agreement with experimental data.

\begin{figure}[h]
\centering
\includegraphics[width=0.40\textwidth, angle=0]{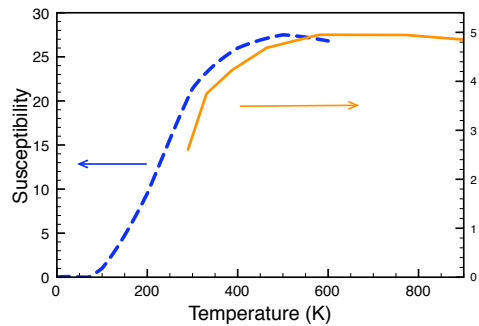}
\caption {(Color online) Spin susceptibility  $\chi(T)$ ( in $\mu_B^2 /$ eV)  from LDA+DMFT calculations (orange solid line) and from experiment \cite{jaccarino} (blue dashed line).}
\label{khi}
\end{figure}

FeSi displays an unusual crossover in the vicinity of the room
temperature from a singlet semiconducting
ground state with a narrow band gap to a
metal with an enhanced spin susceptibility and Curie-Weiss temperature dependence.  \cite{jaccarino}  All previous attempts 
to explain this behavior were based on the models assuming extremely narrow ($<1000$K) peaks at the energy gap edges in DOS, while LDA calculations gave band width nearly an order of magnitude larger than that value. Our DMFT calculations demonstrate that the correlated band insulator model with realistic  DOS can reproduce anomalous temperature dependence of the magnetic susceptibility for FeSi.
  We have computed $\chi(T)$ as a ratio 
\begin{eqnarray}
\chi (T) =  \frac{M}{h},
\end{eqnarray}
where $h$ is a small uniform external magnetic field and $M$ is an induced magnetization of the system. The comparison of the experimental and calculated magnetic susceptibilities is presented in Fig.  \ref{khi}.
The calculated spin susceptibility increases exponentially from T=0 and reaches maximum at 600 K what is in good agreement with the experimental temperature of 500 K.  However, the calculated absolute values of the magnetic susceptibility at $T > 300 K$ disagrees with experimentally observed $\chi(T)$.  Such a disagreement was observed in the previous theoretical investigation. \cite{urasaki} Varying the Coulomb interaction U the authors of Ref. \onlinecite{urasaki} were able to fit a theoretical curve to the experimental one. 

In the present investigation the same model DOS (Fig. \ref{LDA-dos}) and Coulomb interaction parameter $U$ value was used to reproduce successfully band narrowing observed in ARPES, \cite{Klein}  optical conductivity \cite{optics} and  temperature dependence of  magnetic susceptibility. \cite{jaccarino}

\section{$ {\bf Fe_{1-x}Co_{x}Si }$}
\label{FeCo}
The correct theoretical description of magnetic properties for transition metal monosilicides such as for example MnSi and Fe$_{1-x}$Co$_{x}$Si alloys presents a longstanding problem. There has been a considerable amount of experimental and theoretical work
on MnSi and Fe$_{1-x}$Co$_{x}$Si, regarding their structural, magnetic and electronic properties. 
However, at the moment there is no satisfactory first-principles description of monosilicides magnetic properties.
For instance, in case of MnSi it was found that the experimental value of the magnetic moment is about 0.4 $\mu_{B}$. Different first-principles calculations based on the density functional theory give much larger magnetic moment value of 1 $\mu_B$. \cite{jeong,jarlborg}

Fe$_{1-x}$Co$_{x}$Si alloys  are magnetic for almost all of the intermediate 
concentration regimes,\cite{Beille1,Beille2,Manyala} while the end compounds FeSi and CoSi are nonmagnetic, 
the latter being a diamagnetic semimetal.  Fe$_{1-x}$Co$_{x}$Si system is also interesting for scientists due to the promising properties for spintronic device applications.
For instance, in paper \cite{Fisk} the authors have reported the discovery of a large anomalous Hall effect for Fe$_{1-x}$Co$_{x}$Si.
They have demonstrated that the large effect is most likely intrinsic -- derived from the band-structure effects rather than 
due to impurity scattering. They have proposed to consider the transition metal monosilicides as
potential alternatives to the (GaMn)As and (GaMn)N which are the most popular materials for spintronics. 

From theoretical side no calculation reported so far seems to reproduce correctly both magnetic moment value and Curie temperature of  Fe$_{1-x}$Co$_{x}$Si system. To model Fe$_{1-x}$Co$_{x}$Si alloys the authors of paper \cite{Guevara} have used the full potential linearized augmented plane-wave method in combination with a virtual crystal approximation (VCA) as well as  with a supercell approach. The resulting magnetic moments agree with experimental values only for $x < 0.25$.  Having supposed an important role of ordering and segregation they have simulated several alloy configurations for concentrations $x\leq0.5$. The weighted average magnetic moment which was calculated through Boltzmann distribution is in reasonable agreement with experimental values. However, there are no experimental results which support the segregation and ordering phenomena. Moreover,  the results of neutron measurements \cite{Mezei} have demonstrated a random distribution of transition metals in Fe$_{0.5}$Co$_{0.5}$Si.

To simulate randomly distributed Fe$_{1-x}$Co$_{x}$Si alloys the authors of Ref. \onlinecite{Punkkinen} have used the combination of an exact  muffin-tin orbitals method and a coherent potential approximation. They have found an extreme sensitivity of magnetic properties to the internal structure parameters and lattice constant. However, the calculated magnetic moments at concentrations $x>0.3$ still disagreed with those experimentally observed. 

In this paper we have investigated  electronic structure and magnetic properties of Fe$_{1-x}$Co$_{x}$Si alloys 
using statical (LSDA) and dynamical (DMFT) mean-field approaches.
The virtual crystal approximation gives us opportunity to investigate the electronic structure of 
Fe$_{1-x}$Co$_{x}$Si system in whole range of concentrations.
We have found that LSDA results strongly overestimate magnetic moment values and extend magnetic phase diagram to much large values of Co concentration $x$ comparing with experiment. 
An account of correlation effects within DMFT results in good agreement with experiment for magnetic moment values as well as the position of the magnetization $M(x)$ maximum. We came to conclusion that Fe$_{1-x}$Co$_{x}$Si is an itinerant electrons system which magnetic properties can be correctly described by LDA+DMFT method.

\subsection{Statical mean-field results}
\begin{figure}[h]
\centering
\includegraphics[width=0.45\textwidth, angle=0]{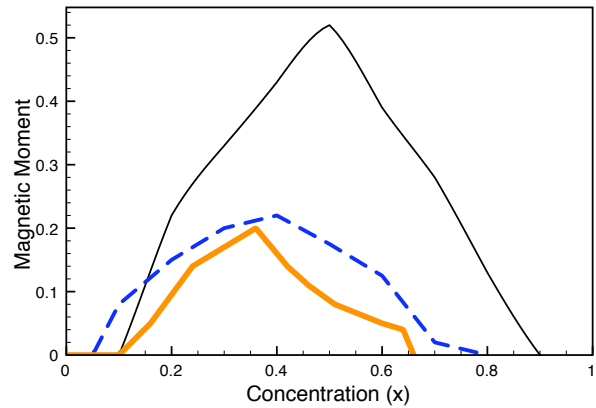}
\caption {(Color online) Concentration dependences of the magnetization (in $\mu_B$) obtained from neutron scattering experiment \cite{Beille2} (blue dashed line), LDA+DMFT (orange bold line) and virtual crystal approximation (black thin line).}
\label{magnetization}
\end{figure}

\begin{figure}[ht]
\centering
\includegraphics[width=0.40\textwidth]{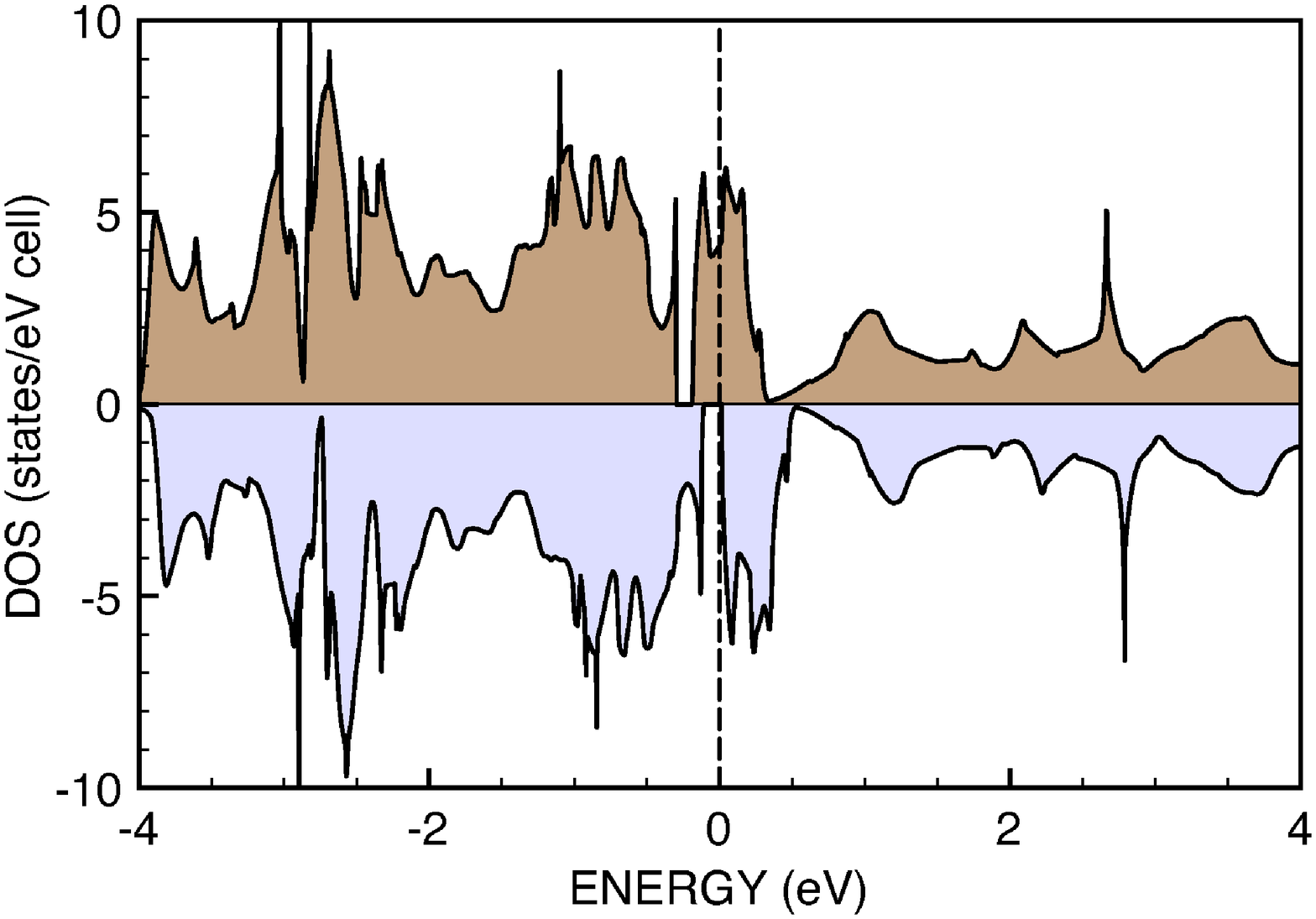}
\includegraphics[width=0.40\textwidth]{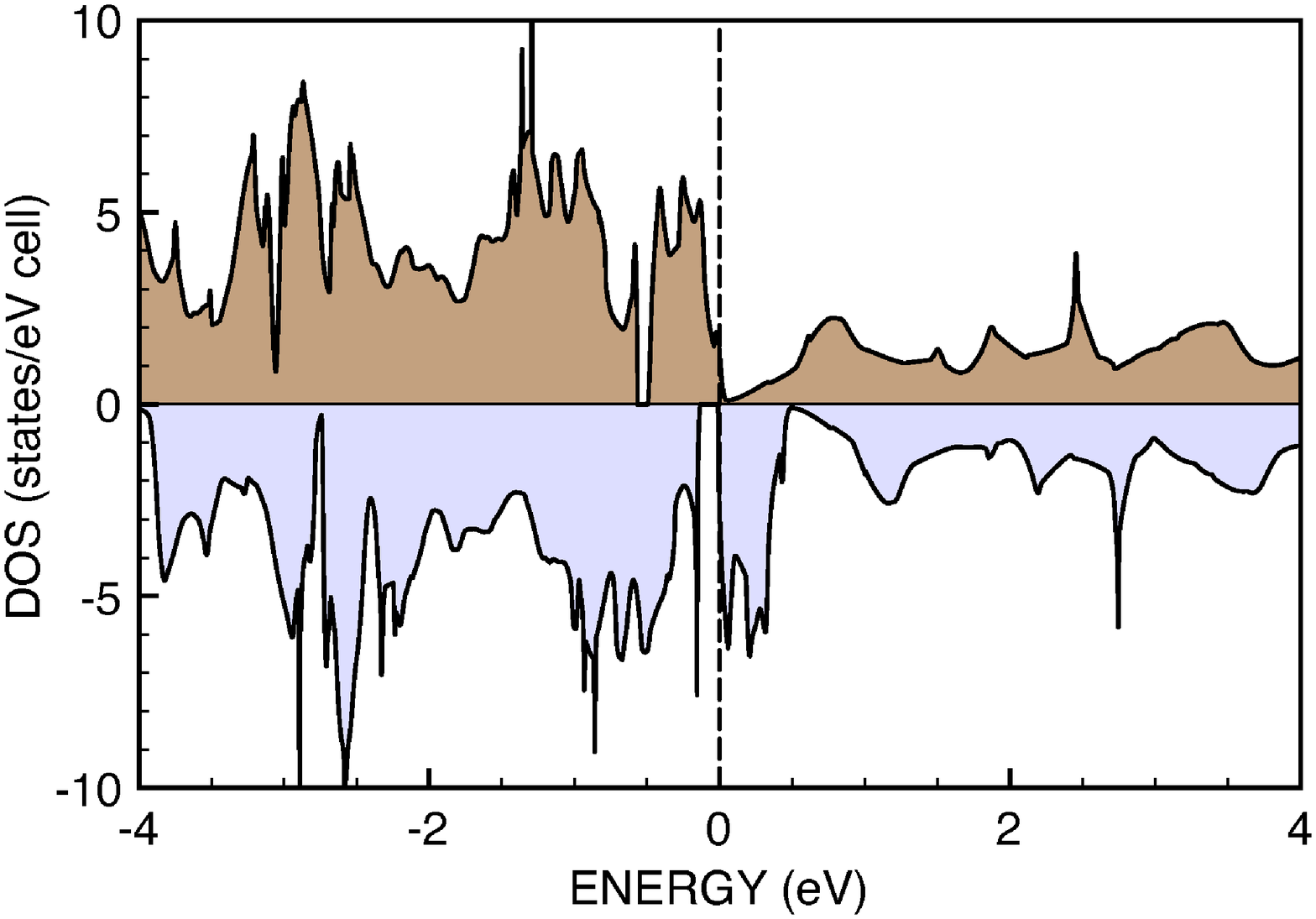}
\includegraphics[width=0.40\textwidth]{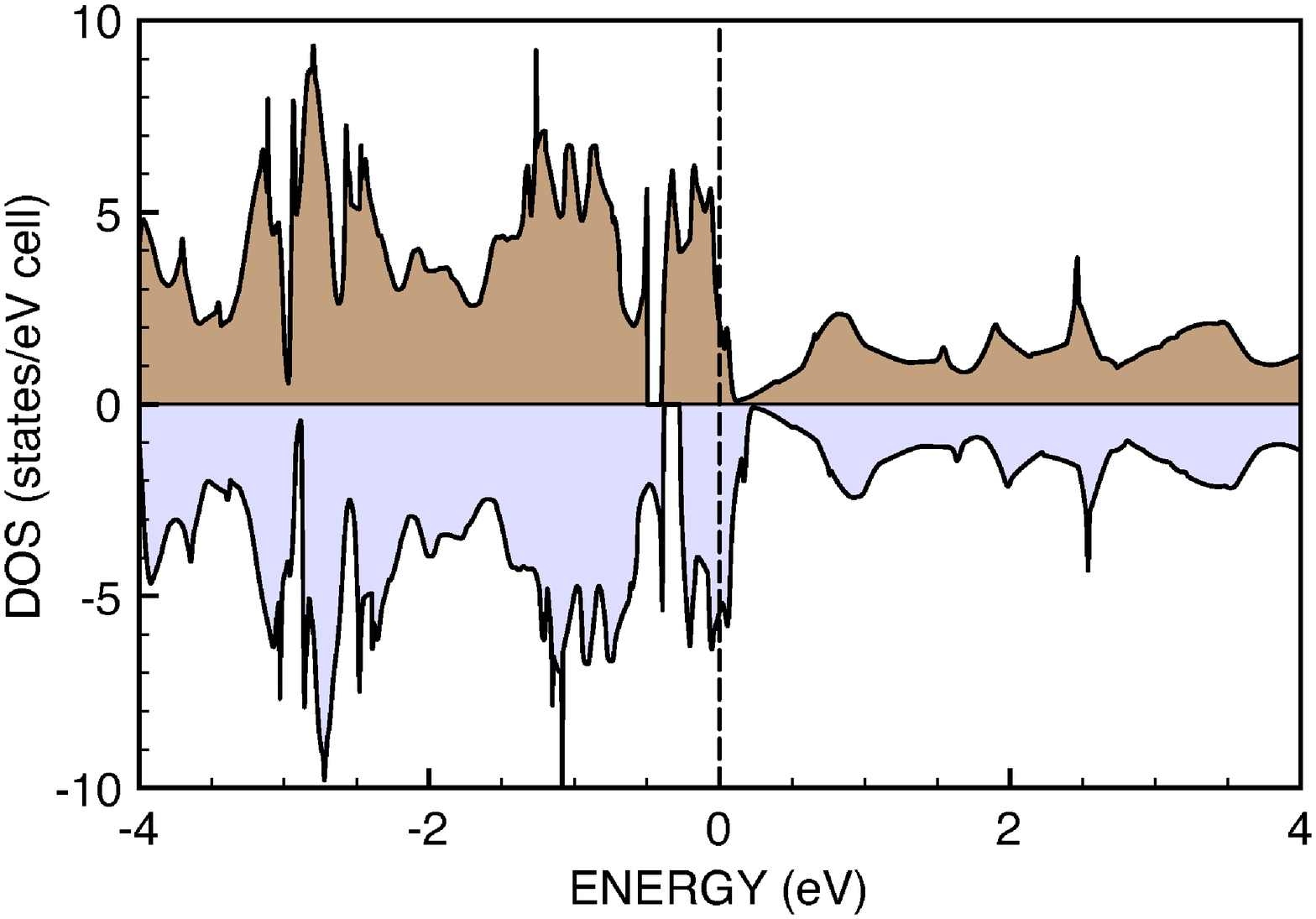}
\caption {(Color online) Density of states for x=0.2, 0.5 and 0.8 obtained in LSDA calculations based on virtual crystal approximation.}
\label{FeCo-LDA}
\end{figure}

The electronic structure of Fe$_{1-x}$Co$_{x}$Si was calculated using the Tight Binding Linear-Muffin-Tin-Orbital 
Atomic Sphere Approximation (TB-LMTO-ASA) method with conventional local-density approximation (LDA).\cite{Andersen} 
The Fe and Co atoms in Fe$_{1-x}$Co$_{x}$Si alloy were treated by virtual atoms with the atomic number value averaged by the
concentration $x$. The experimentally observed lattice constants \cite{Fisk}  were used in the present band structure calculations.
The calculated magnetic moment as a function of cobalt concentration is presented in Fig. \ref{magnetization} and densities of states for x=0.2, 0.5 and 0.8 are presented in  Fig. \ref{FeCo-LDA}. 
One can see that the  calculated results start strongly deviate from experimental values of magnetic moments  for Co concentration  $x>0.3$ with magnetization $M(x)$ maximum at 0.5 $\mu_{B}$ instead of experimental value of about 0.2 $\mu_{B}$ and giving stable magnetism for larger $x$ values than it is observed experimentally. 
Our results agree well with those presented in Ref.\onlinecite{Guevara,Punkkinen}. 

\begin{figure}[ht]
\centering
\includegraphics[width=0.50\textwidth]{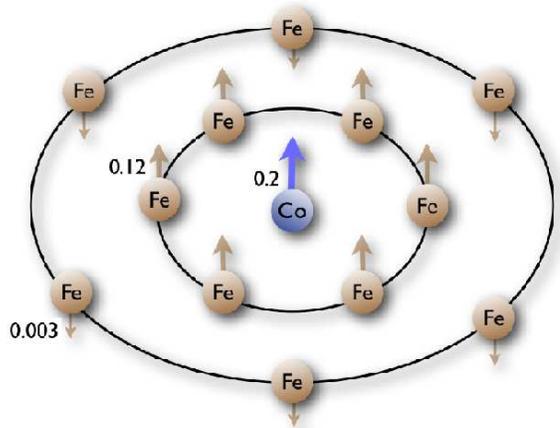}
\caption {(Color online) Schematic representation of a magnetic cluster simulated with LSDA supercell calculations. Numbers are values of magnetic moments of atoms from different coordination spheres. The arrows denote directions of the magnetic moments in the ground state.}
\label{moment}
\end{figure}

A small magnetic moment value calls the question of the nature of magnetism in Fe$_{1-x}$Co$_{x}$Si: do we see itinerant electrons magnetism of the Stoner type or moments are local but their average value is suppressed due to quantum fluctuations and disorder effects?
In order to investigate the localization degree of the magnetic moment in Fe$_{1-x}$Co$_{x}$Si system we have performed the supercell calculation to simulate a Co impurity in FeSi. The supercell was constructed in the FeSi lattice with all basis lattice vectors doubled
and containing a total of 64 atoms of transition metal. The impurity Co atom was assumed to substitute one of Fe atom.
The obtained magnetization spatial distribution in Fe$_{63}$CoSi$_{64}$ is presented in Fig. \ref{moment}.
One can see that magnetic moment is not localized on the Co impurity. There is a magnetic cluster containing one cobalt atom and six nearest iron atoms. This is a result of a strong hybridization of 3$d$-states cobalt and iron through 3$s$- and 3$p$-states of silicon.
Such a magnetic cluster can be considered as a model for magnetic alloy Fe$_{0.84}$Co$_{0.16}$Si for which the experimentally observed magnetic moment 
is about 0.13 $\mu_{B}$ per transition metal atom. The averaged value of the magnetic moment in the cluster of 0.13 $\mu_{B}$ is in excellent agreement with the experimental value. 

The fact that the magnetic moment induced in FeSi by Co alloying is delocalized  supports the itinerant magnetism picture. In this case the Stoner approach is expected to give a realistic  description of Fe$_{1-x}$Co$_{x}$Si magnetic properties. However, the results of this section and previous works show that direct Stoner theory application leads to the overestimation of the magnetic moment value at intermediate concentrations (Fig.\ref{magnetization}).  This is a result of ignoring dynamical correlation effects that, as we will show below, result in a strong renormalization of states near the Fermi level and subsequent reduction of the magnetic moment value.

\subsection{Dynamical mean-field results}
In this section we investigate the influence of dynamical correlation effects on the magnetic properties of Fe$_{1-x}$Co$_{x}$Si.  In our calculations we assume that Coulomb correlations treated by DMFT renormalize a paramagnetic density of states and then this new DOS is used as an input for Stoner theory calculations.

\begin{figure}[ht]
\centering
\includegraphics[width=0.40\textwidth, angle=0]{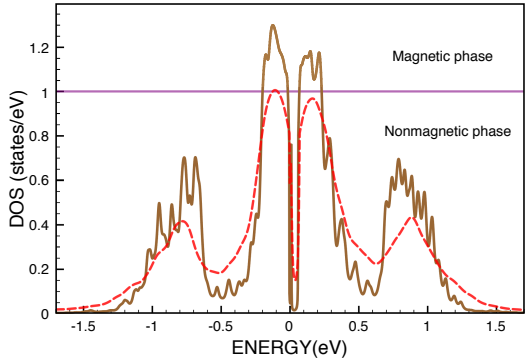}
\caption {(Color online) Density of states for FeSi obtained by using exact diagonalization DMFT method at T=232 K  (red dashed line) and T= 58 K (brown solid line).}
\label{dos-T}
\end{figure}

Let us first qualitatively discuss the origin of magnetism of the investigated alloys using the Stoner criterion 
\begin{eqnarray}
\label{stoner}
I_{d} N(E_{F}) >1.
\end{eqnarray}
Assuming the Stoner parameter $I_{d}$ value is equal to 1 eV the magnetic ground state is stable if the density of states at the Fermi level $N(E_{F})$ is larger than 1 states/eV. With Co substituting Fe in FeSi number of electrons per formula unit increases and the Fermi level runs through the peak above the energy gap. Hence for  Fe$_{1-x}$Co$_{x}$Si the value of  $N(E_{F})$ is determined by the height of this peak. In DMFT the spectral function depends on the temperature.  Fig. \ref{dos-T} gives DMFT densities of states obtained by using the exact diagonalization technique. One can see that the  density of states value at the peak maximum is larger than 1 states/eV at $T$= 58 K but becomes smaller than 1 at  $T$=232 K.  Then the Stoner criterion for magnetism Eq.(\ref{stoner}) is satisfied for  $T$= 58 K but not for $T$=232 K.
This result agrees with experimental values of the Curie temperature ($T_{c}$ = 50 K). 

\begin{figure}[!t]
\centering
\includegraphics[width=0.40\textwidth, angle=0]{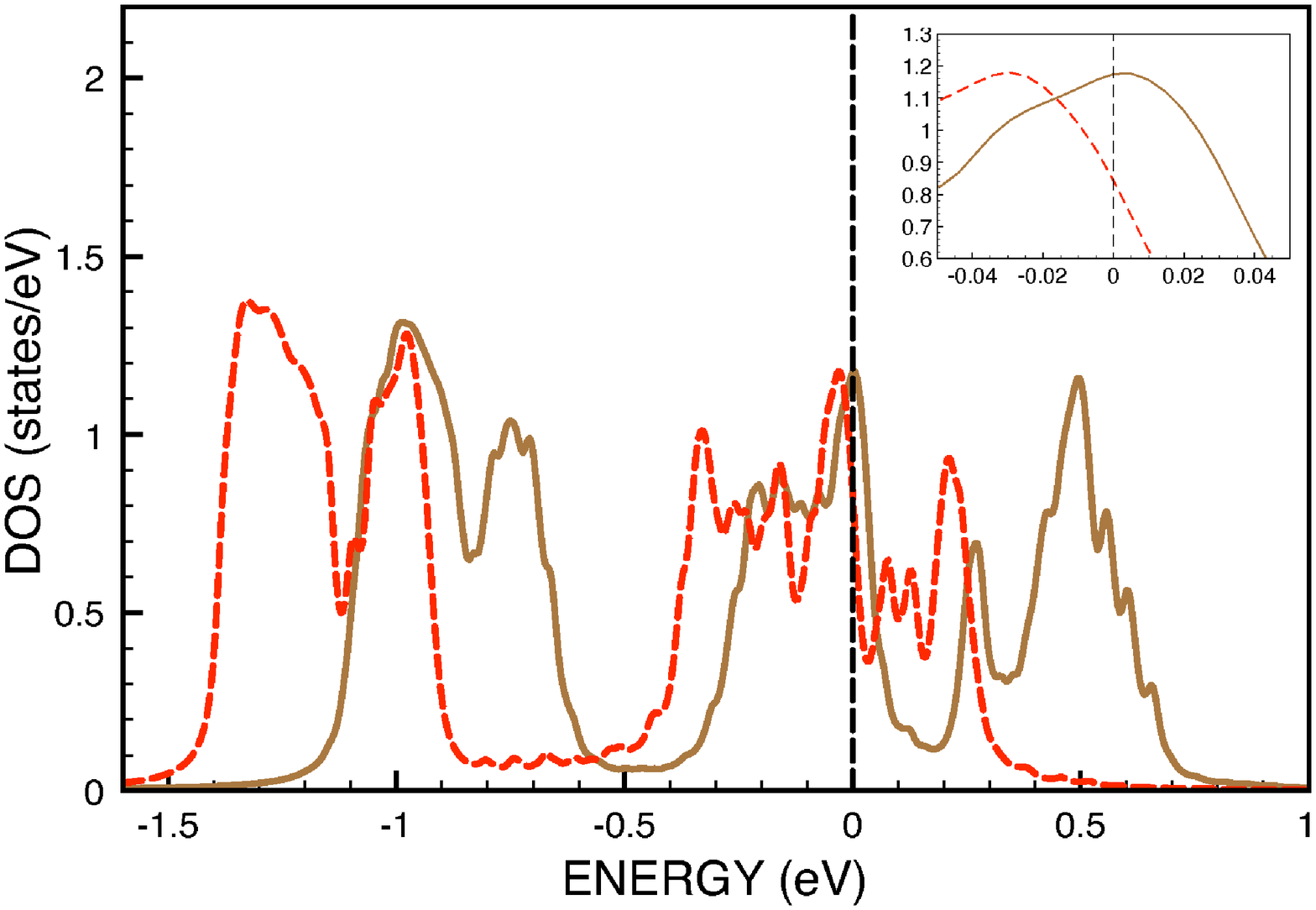}
\caption {(Color online) Density of states of the doped FeSi model obtained by using the exact diagonalization DMFT method at T=58 K  for x=0.36 (brown solid line) and x= 0.66  (red dashed line). The density of states near the Fermi level is presented in the inset. The black dashed line corresponds to the Fermi level.}
\label{dope-dos}
\end{figure}

We are now in a position to perform the quantitative analysis of the magnetism.  The DMFT calculations for  Fe$_{1-x}$Co$_{x}$Si were performed for various  concentrations $x$ values. The results for $x$=0.36 and $x$=0.66 are presented in Fig. \ref{dope-dos}. One can see that for x=0.36 the density of states at the Fermi level $N(E_F)> 1$ and for x=0.66 $N(E_{F}) < 1$. According to the Stoner criterion  Eq.(\ref{stoner}) that gives magnetic and nonmagnetic ground states for $x$=0.36 and $x$=0.66 correspondingly in good agreement with experimental data (see Fig. \ref{magnetization}).
 
We have used the obtained paramagnetic DMFT densities of states for different concentrations (Fig.\ref{dope-dos}) to solve the Stoner model.  
Self-consistent values for 
 spin-up $n_{\uparrow}$ and spin-down $n_{\downarrow}$ numbers of electrons are given by equations for the total magnetic moment 
\begin{eqnarray}
M = \int_{-\infty}^{+\infty} (N(\epsilon + I_{d} n_{\uparrow}) - N (\epsilon + I_{d} n_{\downarrow})) f(\epsilon, T) d\epsilon, 
\end{eqnarray}
and the total number of electrons 
\begin{eqnarray}
N = \int_{-\infty}^{+\infty} (N(\epsilon + I_{d} n_{\uparrow}) + N (\epsilon + I_{d} n_{\downarrow})) f(\epsilon, T) d\epsilon 
\end{eqnarray}
that are recalculated iteratively. Here $N(\epsilon)$ is a density of states obtained in DMFT calculations and $f(\epsilon, T)$ is the Fermi distribution function. The Stoner parameter $I_{d}$ was chosen to be 1 eV close to that used in previous theoretical estimations. \cite{Punkkinen} 

The calculated concentration dependence of magnetization $M(x)$ is presented in Fig. \ref{magnetization}. There is good agreement between
experimental and theoretical values. The main effect of using  $\rho(\epsilon)$ obtained in DMFT is a strong reduction of the resulting magnetic moment values compared to calculations using unrenormalized LDA DOS. This reduction is due to the quasiparticle weight factor $Z\equiv 1/m^*$ appearing in the numerator for Green function expression Eq.(\ref{Z}). Then the integral over quasiparticle band states near the Fermi level is decreased by a factor of $Z$ comparing with unrenormalized LDA values. As $Z\approx 0.5$ in our DMFT calculations that results in corresponding decrease of $M(x)$ values by this factor.

\section{DISCUSSION}
In this paper we have investigated electronic structure and magnetic 
properties of FeSi and Fe$_{1-x}$Co$_{x}$Si systems using statical and dynamical mean-field approaches. Our band structure analysis supports the correlated band insulator model for this materials in contrast to the Kondo insulator model.
The results of DMFT calculations have shown a strong renormalization of states near the Fermi level. The estimated band-mass renormalization $m^*\approx 2$ 
agrees with that obtained in the recent ARPES experiments. 
Analyzing paramagnetic DMFT densities of states calculated at different temperatures and at different Co concentrations we have shown that itinerant magnetism picture is valid for Fe$_{1-x}$Co$_{x}$Si alloys.

\section{ACKNOWLEDGMENTS}
The hospitality of the Institute of Theoretical Physics of Hamburg University (SFB 668) and the Institute of Theoretical Physics of ETH-Zurich is gratefully acknowledged.
We would like to thank T.M. Rice, M. Sigrist, A. L$\ddot{a}$uchli, Y. Yamashita, D. van der Marel, I. Solovyev for helpful discussions and Y.O. Kvashnin for his assistance with Wannier function analysis.
This work is supported by the scientific program ``Development of scientific potential of universities'' N 2.1.1/779, President of Russian Federation 
fund for support for scientific schools NSH 1941.2008.2, 
RFFI 07-02-00041 and RFFI-09-02-00431a, grants of Ural Division of Russian Academy of Science N7 and N28.
The calculations were performed on the computer cluster of ``University Center of Parallel Computing'' of USTU-UPI.

\end{document}